\documentstyle[prl,multicol,aps]{revtex} 
\begin{document}
\draft
\title{Nodes of the superconducting gap probed by electronic Raman  
scattering in $HgBa_{2}CaCu_{2}O_{6+\delta } $ single crystals} 
\author{A. Sacuto}
\address{Physique des Solides, Universit\'e Pierre et Marie Curie,
Case 79, 4 Place Jussieu, 75005 Paris , France}
\author{R. Combescot}
\address{Physique Statistique, Ecole Normale Sup\'erieure,
24 rue Lhomond, 75005 Paris , France}
\author{N. Bontemps, P. Monod}
\address{Physique de la Mati\`ere Condens\'ee, Ecole Normale Sup\'erieure,
24 rue Lhomond, 75005 Paris , France}
\author{V. Viallet and D. Colson}
\address{Physique de l'Etat Condens\'e, DRECAM/SPEC, CEA, Saclay,
91191 Gif sur Yvette , France}
\date{Received \today}
\maketitle
\begin{abstract}
Pure electronic Raman spectra with no phonon structures superimposed  
to the electronic continuum, are reported, in optimally doped  
$HgBa_{2}CaCu_{2}O_{6+\delta } $  single crystals $(T_{c }=126 $ K).  
As a consequence, the spectra  in the $A_{1g }, $ $B_{1g } $  
and $B_{2g } $ symmetries, including the crucial low energy frequency  
dependence of the electronic scattering, are directly and reliably  
measured. The $B_{2g } $ and, most strikingly,
the $B_{1g } $ spectra exhibit a strong intrinsic linear term, which  
suggests that the nodes are shifted from the [110] and [1$\bar{1}$0] 
directions, a result inconsistent with a pure $d_{x^{2}-y^{2}} $ model.
\end{abstract}
\begin{multicols}{2}
\narrowtext
Since the last few years, identifying the symmetry of the pairing  
state has been the expected major step towards an understanding  
of high $T_{c } $ superconductivity. The controversy between s-wave  
or d-wave pairing is not however resolved: some experiments  
appear to advocate strongly in favor of a $d_{x^{2}-y^{2}} $
symmetry \cite{[1]}, whereas others seem to show a significant  
s-wave contribution \cite{[2]}. With respect to this problem, inelastic  
light scattering has been shown to be a powerful  
tool because besides probing the bulk (in contrast with photoemission  
and tunneling), the selection rules on the polarization of incoming  
and outgoing light make the spectra sensitive to the wavevector  
of the electronic excitations \cite{[3],[4],[5],[6]}. 
Usual theoretical approaches 
of electronic Raman scattering ignore other excitations such  
as phonons. The experimental major difficulty lies however in  
the fact that in $La_{2-x }Sr_{x }CuO_{4}, $ $YBa_{2}Cu_{3}O_{7-\delta } $  
 (Y-123), $Bi_{2}Sr_{2}CaCu_{2}O_{8+\delta } $  (Bi-2212) and $Tl_{2}Ba_{2}CuO_{6+\delta } $   
the phonons mask the electronic excitations,
which hampers an accurate determination of their
spectrum in the superconducting  
state  \cite{[7],[8],[9],[10],[11],[12],[13],[14]}.
\par 
In this letter, we report for the first time {\it pure} - {\it  
no phonon structures superimposed} - electronic Raman spectra  
in $HgBa_{2}CaCu_{2}O_{6+\delta } $  (Hg-1212) single crystals.
This compound
belongs to the new mercury family where  
the phonons do not mask the low energy electronic spectrum for  
electric fields within the planes \cite{[15],[16]}.
Its crystallographic  
structure is purely tetragonal $(^{1}D_{4h})\cite{[17]}, $ which allows  
an unambiguous comparison with theoretical calculations based  
on tetragonal symmetry.
Hg-1212 is therefore particularly well suited for the study 
of the superconducting  
gap. Our most striking result is that the low frequency behavior  
of the $B_{1g } $  spectrum displays a quasi-linear dependence;  
this has not been reported sofar, presumably because the determination  
of the spectrum in this frequency range is usually hindered  
by phonons. Such an observation strongly suggests that the nodes  
are shifted from the [110] and [1$\bar{1}$0] directions, 
a conclusion at odds with the generally accepted  
$d_{x^{2}-y^{2}} $ gap symmetry. We believe that our new and unambiguous  
experimental data should be taken into account in any theoretical  
approach of high-$T_{c } $ superconductors.
\par 
Raman spectra have been obtained from high-quality single crystals  
of Hg-1212. The crystals were grown by a single step synthesis  
as previously described for Hg-1223 \cite{[18]}. 
They are perfect parallelepipeds  
with typical 0.5x0.5 mm$^{2} $  cross section and thickness  
0.3 mm. The crystals were characterized by X-ray diffraction  
and energy dispersive X-ray analysis \cite{[18]}. The [100] crystallographic  
direction lies at $45^{o}$ of the edge of the square and the [001]  
direction is normal to the surface. DC magnetic susceptibility  
measurements performed with a SQUID magnetometer yield $T_{c }=126\pm 1K$.
\par  
The measurements were performed with a standard Raman set-up  
using a single channel detection and the $Ar^{+} $ laser 514.52  
nm line \cite{[19]}. The spectral resolution was set at 3 cm$^{-1}. $  
The crystals were mounted in vacuum $(10^{-5} $ mbar) on the  
cold finger of a liquid helium cryostat. The temperature was  
controlled by a Si diode located inside the cold finger. The  
incident laser power density onto the crystal surface was kept  
below 15W/cm$^{2} $  in order to avoid heating of the crystal  
during the low-temperature runs. The incidence angle of the  
incoming light was $60^{o}$ and the scattered light was collected  
along the normal to the crystal surface. The polarizations of  
the incident and scattered lights are denoted in the usual way.  
x: [100] (a axis), y: [010], z: [001] (c axis), x': [110], y':  
[1$\bar{1}$0]. In order to compare our 
experimental data with theoretical  
calculations \cite{[5],[6],[20]}, the pure $A_{1g } $  (xx), $B_{2g } $ (xy)  
and $B_{1g } $ (x'y') symmetries are needed. This requires that  
the incident electric field lie within the xy plane. The  
incident angle being non zero, the crystal must be rotated by  
$45^{o}$ after measuring in the $A_{1g } $ and $B_{2g } $ 
channels in  
order to get the $B_{1g } $ spectrum. The impact of the laser  
beam onto the crystal surface was precisely located 
in order to probe the same crystal area  
for the three symmetries. Finally, a very weakly diffusive spot  
was carefully chosen for each crystal to minimize the  
amount of spurious elastic scattering.
\par 
The raw Raman  
spectra at T=13 K of the Hg-1212 single crystals for the $A_{1g }, $  
$B_{1g } $ and $B_{2g } $ symmetries are displayed in Fig.1a). Remarkably   
and in sharp contrast with Y-123 and Bi-2212 \cite{[9],[13]}, 
no  Raman active  mode hinders the $B_{1g } $ and $B_{2g } $ spectra. 
Weak peaks are seen at 388 and 576 cm$^{-1} $  in the $A_{1g } $  spectrum  
but their energy location is such that no correction is actually  
needed to discuss the electronic spectra \cite{[21]}. Note that  the  
residual  elastic scattering below 50 cm$^{-1} $  is only seen  
in the $A_{1g } $  spectrum and is very weak. Therefore, we are  
in a position to turn immediately to the analysis of the data  
without dealing with the delicate handling of ''phonon subtraction''.  
\par 
Two well marked maxima are observed around 530 cm$^{-1} $ and  
800 cm$^{-1} $  for the $A_{1g } $ and $B_{1g } $  symmetry respectively.   
The ratio of the peak energies is $\sim$ 1.5  
while it was found $\sim$2 for Bi-2212 and Y-123  \cite{[9],[13]}. 
No clear maximum appears in the $B_{2g } $ symmetry, for which  
actually, the scattered intensity levels out smoothly around  530 cm$^{-1}$.
The intrinsic  
(e.g. after subtracting the dark current of the photomultiplier) scattered  
intensities in $A_{1g }, $ $B_{1g } $ and $B_{2g } $ symmetries drop   
close to a zero count rate at zero frequency
whereas, in Bi-2212, a trend towards  
finite intensity was claimed to be observed  
in $A_{1g } $ symmetry \cite{[9]}. We note that in Bi-2212, the presence  
of numerous phonon structures at low frequency especially in  
x'x' polarizations (xx in the notation of ref.\cite{[9]}), makes it  
difficult to estimate quantitatively the electronic scattered intensity.
In this respect, we believe that our  
data are more reliable.
\par 
Fig.1b) displays the imaginary part Im[$\chi (\omega , $  T=13K)] 
of the electronic response functions at T=13 K associated  
to the $A_{1g }, $ $B_{1g } $ and $B_{2g } $ symmetries. They were   
obtained from the raw spectra after subtracting the residual  
Rayleigh scattering, and correcting for the thermal Bose-Einstein  
factor $ 1 + n(\omega) = [ 1 - \exp (-\hbar
\omega / k_{B} T ) ]^{-1} $. 
The dashed lines in Fig.1b) are the imaginary part Im[$\chi (\omega$ ,T=150K)],
of the response functions at T=150 K displayed for  
clarity after smoothing the spectra. Note that the three response  
functions vanish at zero frequency.
The main new observations which emerge from Fig.1b) are: i) the  
$B_{1g } $ response function exhibits a clear decrease of the  
electronic scattering rate at low energy with respect to the  
normal state, unlike the $A_{1g } $ and the $B_{2g } $  response  
functions where the difference is small. ii) the normal state  
intensity is recovered for all symmetries beyond the maximum.  
iii) the two distinct maxima observed for both $A_{1g } $ and $B_{1g } $  
symmetries disappear above T$_{c }$. 
iiii)  the scattered intensity in the $B_{1g } $ channel exhibits  
a quasi-linear increase.
\par 
Our experimental data are of  
special interest in the low frequency regime. Indeed the $B_{1g } $  and  
 $   B_{2g } $    symmetries are the ones which are expected  
to have the most different behavior 
$(\omega ^{3} $ and $\omega  $ respectively)   
in the $d_{x^{2}-y^{2}} $ model.
Two fits have been performed for each symmetry: a  
first one to a power law $\omega ^{\alpha } $  and a second one to the simple  
polynomial function b$\omega$ +c$\omega ^{3}$. 
The $\alpha  $  exponents calculated   
below 300 cm$^{-1}$ for $A_{1g }, $ $B_{2g } $ and $B_{1g } $ symmetries   
are $1.3\pm 0.3, $ $0.8\pm 0.2, $ $1.5\pm 0.5 $ respectively. 
The power law fit   
shows clearly that the low energy range of the $B_{1g } $ spectrum  
does not display a $\omega ^{3} $ dependence, in contrast to the claim  
of ref.\cite{[5],[6],[20]}.
A linear $\omega  $ dependence in the low frequency regime is quite  
compatible with our three response functions. 
In the second fit the ratio $\tau$=b/(c$\omega^{2}$), which gives  
the relative weight of the linear and the cubic variation. 
is $\tau$ = 2$\pm$1, 4$\pm$1, 3.3$\pm$0.7 and 3.4$\pm$0.5 
at 300, 400, 500 and 600 cm$^{-1}$ 
respectively, much larger than an earlier estimate \cite{[14]}.  
This provides quantitative evidence for  
the predominance of the linear part compared to the cubic part  
especially for the $B_{1g } $ symmetry. 
We suspect that the $\omega ^{3} $ fit to  
previous experimental measurements is less reliable 
due to the phonon background \cite{[9],[13]}.
\par 
We now turn to the comparison of our data with existing theories.  
In the limit where the superconducting coherence length is much  
smaller than the optical penetration depth, the imaginary part  
of the unscreened response function for 
$T\rightarrow 0 $ is given by \cite{[3]}:
\begin{eqnarray}
\label{eq1}
\mathrm{Im}[\chi (\omega ,T\rightarrow 0)]={2 \pi N_{F} \over \omega }
\mathrm{Re}{\left\langle{{{\left|{{\gamma}_{k}}\right|}^{2}
{\Delta }_{k}^{2} \over
{({\omega }^{2}-4{\Delta }_{k}^{2})}^{{1 \over 2}}}}
\right\rangle}_{\mathbf{k}}
\end{eqnarray}
$N_{F} $ is the density of states for both spin orientations at  
the Fermi level, and the brackets indicate an average over the  
Fermi surface. $\Delta _{k } $ stands for the superconducting, k-dependent   
gap. $\gamma _{k } $ is the Raman vertex.
\par 
Obviously, our results cannot be accounted for with an isotropic  
gap combined with a cylindrical Fermi surface.
Several $\Delta _{k } $  and $\gamma _{k } $ distributions have been 
proposed to reproduce the electronic excitations.
Taking a Gaussian distribution of $\Delta _{k }$ 's
\cite{[4],[8]}, we obtain  
for the $A_{1g } $  and $B_{1g } $ spectra the fits shown in Fig.1b). 
We have added a $\omega ^{1/2} $ function to the bracketed expression  
at energies greater than the $\Delta _{k } $ values, in order to mimic  
the asymptotic recovery of the normal state behavior.
The low energy part of the  
response function in $B_{1g } $ symmetry is not satisfactorily described  
by this type of calculation and moreover
this numerical fit does not yield any  
information on the gap symmetry.
\par 
We have then attempted to fit our $A_{1g }, $ $B_{1g } $ and $B_{2g } $   
response functions with the calculations of Devereaux et al. for  
a $d_{x^{2}-y^{2}} $ gap \cite{[20]} which were found to describe  
satisfactorily the Bi-2212 results. 
After adjusting the relative energy location of the 
maxima in $A_{1g }$ and $B_{1g } $, we find a fair agreement  
with our data in the 
linear dependence of the low energy part of the $A_{1g } $ and  
$B_{2g } $ symmetries (see Fig.2). However, the low frequency  
behavior of the theoretical $B_{1g } $ spectrum is incompatible  
with our experimental results. This was already clear from the
above discussion of the exponents and of the ratio of the linear  
and cubic contributions. We have also done the calculation for  
the simplest s-wave anisotropic gap $\Delta (\theta )=
\Delta _{0}+\Delta _{1} \cos^{2} 2\theta , $   
not finding a satisfactory agreement because no hint of a threshold  
corresponding to the minimum gap is seen in our data.
\par 
Let us therefore consider the implications of our experimental  
findings. We analyze first the $B_{1g } $ and $B_{2g } $ symmetries,  
which are easier to interpret due to simple symmetry considerations
and are not affected   
by screening in contrast with the $A_{1g } $ symmetry.
In the $B_{1g }$ case,
the Raman vertex $\gamma _{k } $  
is zero by symmetry for $k_{x }=k_{y }, $   
hence the electronic scattering is insensitive to the gap structure  
around $45^{o}$ ; in contrast, the $k_{x } $ $\approx  $ 0 
and $k_{y } $  $\approx  $ 0 regions  
do contribute, giving weight to the gap $\Delta _{0} $  in these directions.  
Conversely, in the $B_{2g }$ case, the Raman vertex  
$\gamma _{k } $ is zero by symmetry for $k_{x }=0 $ or $k_{y }=0, $ and 
non zero elsewhere, hence provides weight in the $k_{x }=k_{y } $   
direction. In the $d_{x^{2}-y^{2}} $ pairing state, these considerations  
imply that the $B_{1g } $  symmetry is insensitive to the nodes  
at $45^{o}$ (hence the $\omega ^{3} $  dependence) and displays a maximum  
at $2\Delta _{0}, $ whereas the $B_{2g } $ symmetry exhibits a linear  
frequency dependence (because it probes the nodes) and a smeared  
gap. An inescapable consequence of the linear low frequency  
dependence is the existence of a density of states which increases  
linearly with frequency, most naturally due to nodes in the  
gap. Since these nodes are probed in both $B_{1g } $ and $B_{2g } $  
symmetries, \emph{they cannot be (only) located in the  
$45^{o}$ direction}.
\par 
It could be argued that impurities are responsible for the  
observed low frequency density of states in the $B_{1g } $  channel.  
Devereaux \cite{[22]} has shown that, for a $d_{x^{2}-y^{2}} $  
gap, impurities induce for $B_{1g } $ a linear rise of the electronic  
scattering at low frequency, crossing over to the $\omega ^{3} $  dependence  
at higher frequency. However the linear behavior in $B_{1g } $  
extends in our experiment up to 600 cm$^{-1}, $ yielding a cross-over  
energy $\omega ^{*}   \sim (\Gamma \Delta )^{1/2} $  of order 600 cm$^{-1}$.
In the most favorable case of unitary scatterers \cite{[22]} , this would  
imply a scattering rate $\Gamma  $  of order of the gap $\Delta  $ itself.   
Such a high scattering rate should strongly reduce the critical  
temperature, in contradiction with the optimal 126 K critical  
temperature of our samples. Therefore this explanation looks  
unlikely. We believe that our data are representative  
of a pure compound and that the linear contribution found in  
the $B_{1g } $ spectrum is intrinsic.
\par 
We are thus left with the conclusion that the nodes are shifted  
from the $45^{o}$ direction. Hg-1212 being tetragonal, we cannot  
ascribe this to an orthorhombic distorsion as could be the case  
in Y-123. We therefore explore this shift more quantitatively.  
As a first approach, we take the simple model  
$\gamma _{B1g }  \sim \cos 2\theta  $ ,
$\gamma _{B2g }   \sim \sin 2\theta  $ for the Raman  
vertices \cite{[5],[6]} and a cylindrical Fermi surface. We start by  
using a ''toy model'' where we compute the $B_{1g } $ and $B_{2g } $  
spectra for an order parameter 
$\Delta (\theta ) =  \Delta _{0} \cos(2\theta -2\alpha ) $   
obtained by artificially rotating the $d_{x^{2}-y^{2}} $
one by an angle $\alpha . $ This is intended to check the sensitivity  
of the Raman spectra to the nodes location (note that a $22.5^{o}$
rotation would naturally make $B_{1g } $ and $B_{2g } $ identical).  
We find that a rotation by $\alpha  $ $\approx  $$10^{o}$ 
is barely noticeable in  
the spectra : the linear rise at low $\omega  $ produced in $B_{1g } $  
by this rotation is comparable to the experimental accuracy.  
This suggests that the shift of the nodes away from $45^{o}$ is quite  
sizeable.
\par 
Because of the tetragonal symmetry, one node located at $\theta _{0} $  
( $\not=  $ $45^{o}$ or $0^{o}$ ) implies 8 nodes at 
$\pm \theta _{0}+n\pi /2 $ (n = 0,1,2,3   
). A very simple model corresponding to this situation 
is  $\Delta (\theta ) $  
$= $  $\Delta _{0} $  $[\cos $ $(4\theta )$ + s] 
with 0 $\leq  $ s $\leq  $ 1, an order parameter  
which has the $A_{1g } $ symmetry \cite{[23]}. The gap has a maximum  
$\Delta _{0}(1+ $ s) for $\theta =0, $ seen in $B_{1g }, $ and a secondary   
maximum $\Delta _{0} $  $(1-s) $ for $\theta  $ = $\pi /4, $ 
seen in $B_{2g }, $ while  
the node lies at $\theta _{0}=(1/4) $ arccos(-s). This model could  
also be loosely called a g-wave model or a (super) extended  
s-wave model (in contrast to the d-wave model, which has the  
$B_{1g } $ symmetry). With one more parameter we might decouple  
the position of the nodes and the size of the gap maxima. We  
choose the parameter  s  to account for the relative  
peak positions in $B_{1g } $ and $B_{2g } $ ( inasmuch as we consider  
that $B_{2g } $ has a very broad peak around 500 cm$^{-1} $  ).  
It can be seen in Fig.2 that the $B_{1g } $ and $B_{2g } $ spectra  
calculated in this framework for s=0.2 (giving $\theta _{0} $  = $25^{o}$) are  
in good agreement with experiment for the low frequency behavior.  
The discrepancy at high frequency, most conspicuous in $B_{2g }, $ 
is assigned to finite electronic scattering rate in the normal state.
This should be also accounted for, but this is beyond the scope of this letter. 
In order to remove the singularities at the gap maxima, a smearing  
function with width proportional to frequency \cite{[22]} has been  
incorporated in the calculation.
\par 
Our model explains the different positions of the $B_{1g } $  
and $B_{2g } $ peaks (also found in Bi-2212 ) by the gap anisotropy.  
This leads us to try and connect the position of the peak found  
in $A_{1g } $ symmetry to the other peaks.
The $A_{1g } $  symmetry is screened
and the peak position can not be related directly to $\Delta (\theta )$. 
Since a constant vertex is completely screened, the simplest model  
giving a nonzero result for $A_{1g } $  is $\gamma _{A1g } 
\sim \cos 4\theta  $. 
However, in our one  
parameter model, the position of the nodes is imposed by the  
relative peak position in $B_{1g } $ and $B_{2g } $ .
This leads to the $\gamma _{A1g } \sim \cos 4\theta  $  being essentially   
zero at the nodes, and yields a very small low $\omega  $ linear contribution  
in contrast with experiment. We have thus included higher  
harmonics. The calculation shown in Fig.2 is done with $\gamma _{A1g } 
\sim \cos 4\theta - 3 \cos 8\theta $ . The agreement is quite
satisfactory   
both with respect to the low $\omega  $ behaviour and the peak position.  
However it is clear that this could even be improved in a more  
complicated two parameter model for $\Delta (\theta ), $ decoupling the  
node position and the gap maxima, and also including a  
$\theta  $ dependence for the density of states.
\par 
In conclusion, we have presented for the first time, pure  
electronic Raman spectra of Hg-1212 single crystals. Our most  
significant result is the observation of an intrinsic linear  
$\omega  $ dependence -not only in the $B_{2g } $ spectrum- but 
also in the  
$B_{1g } $ spectrum. This is an important clue in order to locate  
the nodes, which was not reported sofar in other compounds presumably  
because the experimental determination of the low frequency  
regime is usually hampered by phonons. Our observations advocate  
in favor of a shift of the nodes from the [110] and [1$\bar{1}$0]
directions, which is inconsistent with the simple $d_{x^{2}-y^{2}} $
model.
\par 
We thank M. Cyrot and E.Ya. Sherman for very fruitful discussions.

\begin{figure}
\caption{ a) Raw Raman spectra of the Hg-1212 single crystals in  
$A_{1g }, $ $B_{1g } $ and $B_{2g } $ symmetries at T=13 K 
b) Imaginary parts of the response function Im$[\chi (\omega ,T=13K)] $  
in the $A_{1g }, $ $B_{1g } $ and $B_{2g }. $ The dashed lines represent   
Im$[\chi (\omega ,T=150K)]. $ The solid line is a fit to the $A_{1g } $ and   
$B_{1g } $ spectra of a Gaussian distribution of gaps 
with $2\Delta _{0}=360 $ cm$^{-1} $  and $\sigma =200 $ cm$^{-1} $ 
for $A_{1g } $   
and $2\Delta _{0}$=670 cm$^{-1} $ and $\sigma$ =180 cm$^{-1} $  
for $B_{1g } $   
$ (\Delta _{0} $ and $\sigma  $ are the mean and the standard deviation  
of the Gaussian respectively)}
\label{Fig1}
\end{figure}
\begin{figure}
\caption{From left to right: $A_{1g}, $ $B_{1g} $ and $B_{2g} $  data  
and calculations for the $d_{x 2-y 2} $ model (dashed line)  
and for a model with 8 nodes lying at $\theta _{0}=25^{o}$ (full line)}
\label{Fig2}
\end{figure}
\end{multicols}

\begin{references}
\bibitem{[21]} The 576 $cm^{-1} $  peak shows a phonon asymmetric lineshape  
with an antiresonance on the high energy side. This is characteristic  
of the interference between an electronic continuum and a single  
phonon state. Such Fano lineshape has already been detected  
in Y-123 and Bi-2212 for the $B_{1g } $ normal mode  
in the vicinity of the gap structure maximum [7,10].
\bibitem{[23]} An order parameter with $B_{1g } $
 symmetry (changing sign under  
$\pi /2 $ rotation) would imply 12 nodes.
\end{references}
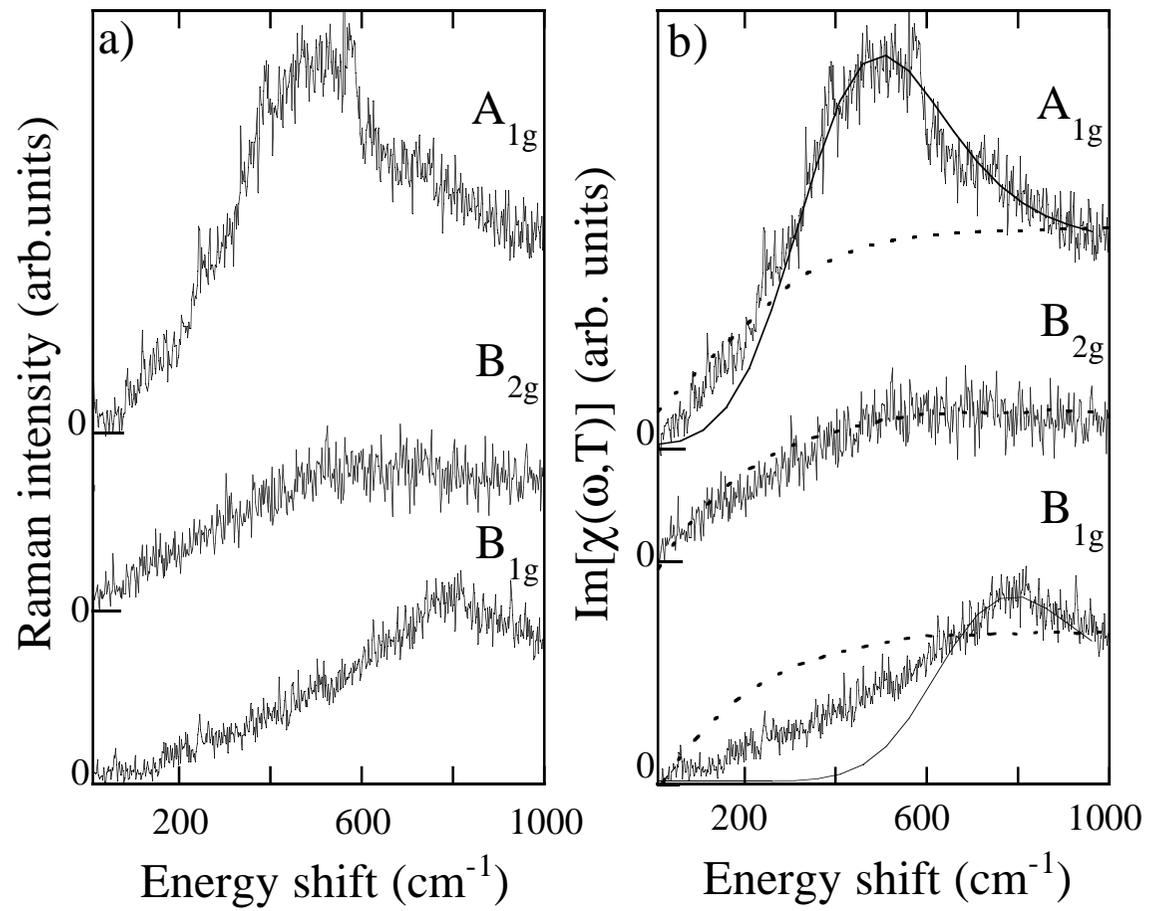
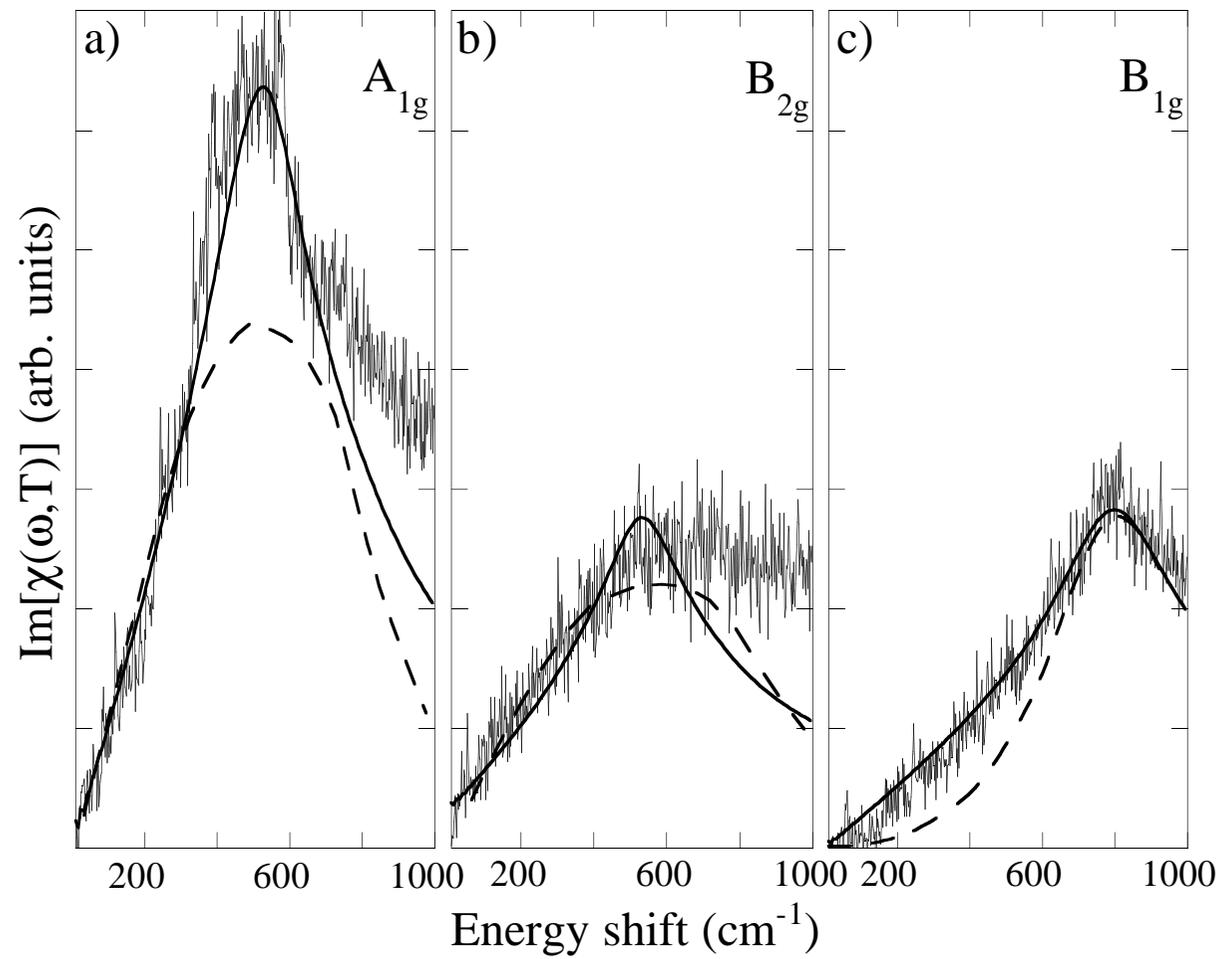
\end{document}